\documentclass[twocolumn,showpacs,preprintnumbers,amsmath,amssymb]{revtex4}

\usepackage{graphicx}
\usepackage{dcolumn}
\usepackage{bm}

\begin{document}

\preprint{ACT-08-10, MIFP 09-38}

\title{On Crossing Symmmetry and Modular Invariance in Conformal Field Theory and S Duality in Gauge Theory}

\author{Dimitri V. Nanopoulos$^{1,2,3}$}
\author{Dan Xie$^1$}
\affiliation{$^{1}$George P. and Cynthia W.Mitchell Institute for
Fundamental Physics,\\ Texas A\&M University, College Station, TX
77843, USA \\ $^{2}$Astroparticle physics Group, Houston Advanced
Research
Center (HARC), Mitchell Campus, Woodlands, TX 77381, USA\\
$^{3}$Academy of Athens, Division of Nature Sciences,\\ 28
Panepistimiou Avenue, Athens 10679, Greece
 }%

\date{\today}

\begin{abstract}
In this note, we explore the relation between crossing symmetry and
modular invariance in conformal field theory and S-duality in gauge
theory. It is shown that partition functions of different S dual
theories of $N=2$ $SU(2)$ gauge theory with four fundamentals can be
derived from the crossing symmetry of the Liouville four point
function. We also show that the partition function of $N=4$ $SU(2)$
gauge theory can be derived from the Liouville partition function on
torus.

\end{abstract}

\maketitle

\section{Introduction}
Historically, string theory is originated as an attempt to provide a
theory of strong interactions. The famous Veneziano amplitude is
shown to be derivable from the vibrating string which sweeps out a
two dimensional world sheet in space time. The scattering amplitude
is calculated from the correlation function on the Riemann surface
which can be regarded geometrically as punctured Riemann surface.
Two dimensional conformal field theory (CFT) plays a central role in
the development of the string theory. The most important dynamical
principle of this approach \cite{polyakov} is the associativity of
the operator product expansion (OPE) which in turn leads to the
crossing symmetry of the four particle scattering amplitude. This
crossing symmetry is the underlying reason to eliminate the high
energy divergences of tree level diagrams. On the other hand,
modular invariance of the one-loop partition function of conformal
CFT leads a miraculous way to eliminate high energy divergences of
one loop diagrams in string theory.

Furthermore, people were trying to understand the strong coupled
dynamics of four dimensional quantum field theory from the
Srong-Weak duality (S dualtiy) of the quantum field theory itself
\cite{gno,mo}. During the past two decades, we gained a lot of
understandings about four dimensional gauge theories from the S
dualities \cite{witten4}. S dualities are also related to Geometric
Langlands Duality \cite{witten1}.

These two seemingly unrelated subjects are put together by Gaiotto's
observation \cite{Gaiotto,Argy} on $N=2$ superconformal field theory
(SCFT)(see recent developments in (\cite{Gaiotto2,yuj1,
morozov,seji,mori,dan1,Gaiotto5}). It is shown that four dimensional
$N=2$ SCFT can be realized as six dimensional $(0,2)$ theory
compactified on a punctured Riemann surface. The coupling constants
are determined by the complex structure of this punctured Riemann
surface while different S dual frames are determined by various
degeneration limits of this surface. The flavor groups are
determined by the puncture types which are labeled by Young
Tableaux. Everything about the gauge theory is encoded into this
punctured Riemann surface. Recalling our discussion about the
conformal field theory, it is natural to wonder if there is any
connection between those two subjects.

AGT \cite{Gaiotto3} proposed a remarkable conjecture that for
$SU(2)$ theory the partition function of the gauge theory is
equivalent to the correlation function of the Liouville theory.
Liouville theory falls into the general framework of CFT \cite{zz}.
One may wonder why a four dimensional theory can be equivalent to a
two dimensional theory. This might be related to a remarkable
property of the Liouville theory: the central charge is adjustable.
Then the degrees of freedom of Liouville theory is rather mysterious
and makes a four dimensional correspondence possible.

It is interesting to further explore the correspondence between
gauge theory and Liouville theory. In this note, we make the
observation that we can determine the transformation law under
S-duality of the partition function of $N=2$ gauge theory with four
fundamentals from the crossing symmetry of the Liouville four point
function. The transformation law for $N=4$ $SU(2)$ theory under
S-duality can be determined similarly from the one-loop partition
function of Liouville theory.

\section{Crossing Symmetry and Modular Invariance in Conformal Field Theory}
Let's first review how the dual string model was proposed
\cite{gsw}. Consider an elastic scattering amplitude with incoming
spinless particles of momentum $p_1$, $p_2$ and outgoing spinless
particles of momentum of $p_3$, $p_4$ (see Figure 1). The
conventional Mandelstam variables are
\begin{equation}
s=-(p_1+p_2)^2,~~t=-(p_2+p_3)^2,~~u=-(p_1+p_4)^2.
\end{equation}

\begin{figure}
\begin{center}
\includegraphics[width=3.5in,]
{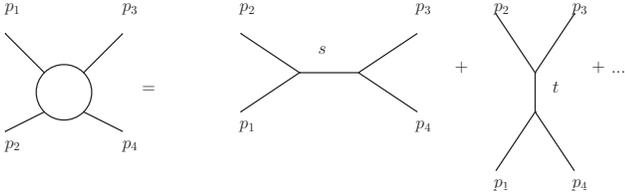}
\end{center}
\caption{An elastic scattering with incoming particles of momentum
$p_1$, $p_2$ and outgoing particles of momentum $p_3$, $p_4$. We
indicate the contribution from $s$ channel and $t$ channel. The
field theory amplitude is constructed from the sum of those
contributions. }
\end{figure}

Consider first the $t$ channel contribution. There are various
particles with mass $M_J$ and spin $J$ which might be exchanged:
\begin{equation}
A(s,t)=-\sum_J{g_J^2(-s)^J\over t-M_J^2}.
\end{equation}
We have the following similar amplitude if we consider $s$ channel:
\begin{equation}
A^{'}(s,t)=-\sum_J{g_J^2(-s)^J\over s-M_J^2}.
\end{equation}
Two remarkable properties of the scattering amplitude are that the
above sums are infinite and these two amplitudes are equal to each
other $A(s,t)=A^{'}(s,t)$. The last property which is called $s-t$
duality motivates the proposed Veneziano amplitude.

It is well known that the Veneziano amplitude can be derived from
two dimensional (world sheet) string theory. The infinite sum is due
to the infinite number of states in mass spectrum. The $s-t$ duality
is simply the crossing symmetry of the four point function of
conformal field theory. This crossing symmetry is also equivalent to
the associativity of the OPE on the world sheet:
\begin{equation}
A_i(\zeta)A_j(0)=\sum_kC_{ij}^k(\zeta)A_k(0).
\end{equation}

We briefly summarize some properties of CFT. The Virasoro algebra is
\begin{equation}
[L_n,L_m]=L_{m+n}+{c\over12}n(n^2-1)\delta_{n+m,0},
\end{equation}
here $c$ is the central charge and $L_m$ are generators of conformal
symmetry. The representations of this algebra are labeled by primary
states which satisfy:
\begin{equation}
L_0|V_\alpha>=\Delta_\alpha |V_\alpha>,~~~ L_n |V_\alpha>=0,~~n>0,
\end{equation}
$\Delta_\alpha$ is the conformal dimension of this primary state.
The other states of this representation are represented as:
\begin{equation}
L_{-k_n}L_{-k_{n-1}}....L_{-k_1}|V_{\alpha}>,
\end{equation}
here $k_n\geq k_{n-1}...\geq{k_1}$. These secondary states have
conformal weights $\Delta=\Delta_\alpha+|Y|$; here $|Y|$ is the
total boxes of the Young Tableaux with rows $k_1,...k_n$. The
correlation functions involving the energy momentum tensor and
secondary states are expressed by the correlation functions of the
primary states.

The OPE of two primary states are given as \cite{polyakov}:
\begin{eqnarray}
\phi_m(z,\bar{z})\phi_n(0,0)=\sum_p
c_{nm}^pz^{\Delta_p-\Delta_n-\Delta_m}\nonumber\\
\bar{z}^{\bar{\Delta}_p-\bar{\Delta}_n-\bar{\Delta}_m}\psi_p(z,\bar{z}|0,0).
\end{eqnarray}
The most important dynamical information are $c_{nm}^p$ and the
conformal dimensions.

The four point function has the form
\begin{equation}
G^{lk}_{nm}(x,\bar{x})=<k|\phi_l(1,1)\phi_n(x,\bar{x})|m>,
\end{equation}
here we fixed the positions of three vertex operators as
$0,1,\infty$ and $x$ is the projective invariant variable. The
crossing symmetry is
\begin{equation}
G_{nm}^{kl}(x,\bar{x})=G_{nl}^{mk}(1-x,1-\bar{x})=x^{-2\Delta_n}\bar{x}^{-2\bar{\Delta}_n}
G_{nk}^{lm}({1\over x}, {1\over \bar{x}}) \label{1}
\end{equation}
Using OPE of $\phi_n\phi_m$, the four point function is
\begin{equation}
G^{lk}_{mn}(x,\bar{x})=\sum_pc_{nm}^pc_{lkp}F_{nm}^{lk}(p|x)\bar{F}_{nm}^{lk}(p,\bar{x}),
\end{equation}
here $F$ is the conformal block which is entirely determined by the
conformal symmetry(we give the s channel contribution). The crossing
symmetry relates the contributions of different channels (see for
instance Figure 2):
\begin{figure}
\begin{center}
\includegraphics[width=3.5in,]
{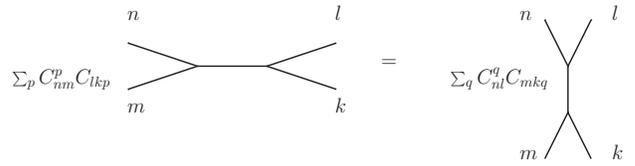}
\end{center}
\caption{Crossing symmetry of four point function in conformal field
theory. }
\end{figure}

Next let's consider the one-loop partition function defined on a
torus:
\begin{equation}
Z(\tau)=\sum_i q^{h_i-c/24}\bar{q}^{\tilde{h}_i-c}(-1)^{F_i},
\label{3}
\end{equation}
here $i$ runs over all the states in CFT and $F_i$ is the fermion
number, $q=\exp^(2\pi i\tau)$. It is well known that this partition
function is needed to be invariant under the $SL(2,Z)$ modular group
transformation of the torus. The high energy density states of this
theory is determined by the central charge for compact unitary
conformal field theory. For Liouville theory, there is a
modification to this result, see \cite{dan}.

\section{Superconformal Field Theories in Four Dimensions}
It is proposed in \cite{mo,gno} that a strongly coupled gauge theory
can be described by a weakly coupled theory in which the elementary
particles are the monoples of the original theory. This proposal is
naturally realized in $N=4$ supersymmetric gauge theory and is
extended to a $SL(2,Z)$ duality group. Later, $D=4$ $N=2$ $SU(2)$
gauge theory with four fundamental hypermultiplets has also been
shown to have the $SL(2,Z)$ duality \cite{witten4}. Gaiotto found an
extremely useful way to describe the gauge structure of different
S-dual frames of this theory. We use the $SU(2)_a\bigotimes
SU(2)_b\bigotimes SU(2)_c\bigotimes SU(2)_d$ subgroup of full
$SO(8)$ flavor group. This theory can be realized as the six
dimensional $(0,2)$ $A_1$ theory compactified on a sphere with four
punctures. The different S dual frames are realized as the different
degeneration limits of this punctured sphere, see Figure 3.
\begin{figure}
\begin{center}
\includegraphics[width=3.5in,]
{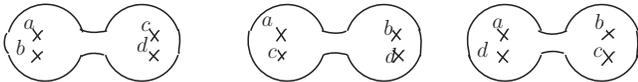}
\end{center}
\caption{Different degeneration limits of the punctured sphere, this
corresponds to different weakly coupled S-dual theory of $N=2$
$SU(2)$ gauge theory with four fundamentals.}
\end{figure}
The Seiberg-Witten (SW) curve of this theory is written as
\begin{equation}
x^2={u\over(z-1)(z-q)z},
\end{equation}
we have fixed the positions of three punctures and left an unfixed
puncture which is identified with the gauge coupling constant. The
SW curve for the mass deformed theory is $x^2=\phi_2(z)$, where
\begin{equation}
\phi_2(z)={M_2(z)\over z^2(z-1)^2(z-q)^2}+{U_2(z)\over z(z-1)(z-q)}
\label{5}.
\end{equation}

For each puncture, we associate a mass parameter $m_{a,b,c,d}$ to
it; the physical masses of the fundamentals are given by
\begin{equation}
m_{1,2}=m_a\pm m_b,~~~m_{3,4}=m_c\pm m_d.
\end{equation}

The partition function of this theory on $S^4$ is given by
\cite{pestun}:
\begin{equation}
Z_{s^4}={1\over vol(G)}\int[da]e^{-{4\pi^2r^2\over
g_{YM}^2}(a,a)}Z_{1loop}|Z_{inst}^N(r^{-1},r^{-1},a)|^2 \label{2}
\end{equation}
Here $a$ is the parameter for the Coulomb branch and $r$ is the
radius of the sphere. The 1-loop part is given by:
\begin{equation}
Z_{1-loop}^{N=2}={H(2a)H(-2a)\over \sum_{i=1}^4H(a+m_i)H(a-m_i)}.
\end{equation}
Here $H(x)$ is given by Barnes's G function\cite{barnes}
$H(x)=G(1+x)G(1-x)$; $m_i$ is the mass parameters for the four
fundamentals and $a$ is the Coulomb branch parameter. The instanton
part of the partition function is identified with the Nekrasov
instanton partition function \cite{nekrasov}
$Z_{inst}(\epsilon_1,\epsilon_2, a)$. Notice that for the $S^4$ case
$\epsilon_1=\epsilon_2={1\over r}$.

Next, let's study the $N=4$ SU(2) theory. It is given by the six
dimensional $A_1$ theory compactified on a smooth torus. The
$SL(2,Z)$ duality of the gauge theory is interpreted as the
$SL(2,Z)$ modular invariance of this torus. The full partition
function of this theory is of the same form as formula (\ref{2}). It
is interesting to note that for $N=4$ $U(M)$ gauge theory the
one-loop part is trivial $Z_{1-loop}=1$ and the full partition
function for $U(M)$ gauge theory is:
\begin{equation}
Z=C{1\over |\eta(\tau)|^{2M}(2\pi\sqrt{\tau_2})^M} \label{4}.
\end{equation}
where $\eta(\tau)=q^{1\over24}\prod_{k=1}^\infty(1-q^k)$, $q=e^{2\pi
i\tau}$.

\section{S Duality from the Conformal Field Theory}
AGT \cite{Gaiotto3} made a conjecture that the full partition
function of the above $N=2$ $SU(2)$ SCFT is equivalent to the
correlation function of Liouville theory. It is shown in
\cite{Gaiotto3} that the instanton part of the gauge theory
partition function is equal to the conformal block of the
correlation function and the one-loop part and classical part
correspond to the structure constant part of the correlation
function. It is also argued that the energy momentum tensor of
Liouville theory is related to the operator (\ref{5}) (this can be
seen from the classical uniformization problem with the punctured
sphere).

The relation between the deformation parameters and the parameters
in Liouville field theory is
\begin{equation}
\epsilon_1=b,~~\epsilon_2={1\over b},
\end{equation}
here $\epsilon_1$ and $\epsilon_2$ are the deformation parameters in
Nekrasov's instanton partition function. Notice that in order to use
the partition function on $S^4$, we need to set $b=1$.

We associate a exponential vertex operator $e^{\alpha_i\phi}$ to
each puncture. We also associate a intermediate state
$e^{\alpha\phi}$ to weakly coupled $SU(2)$ group with Coulomb
parameter $a$. The exact relations between the parameters in gauge
theory and Liouville theory are
\begin{equation}
\alpha_1=m_a+{Q\over2}, \alpha_2=m_b, \alpha=a+{Q\over2},
\nonumber\\
\alpha_3=m_c+{Q\over2}, \alpha_4=m_d.
\end{equation}
Here $Q$ is the conventional parameter for the Liouville theory
$Q=b+{1\over b}$. See Figure 4 for the correspondence.
\begin{figure}
\begin{center}
\includegraphics[width=3.5in,]
{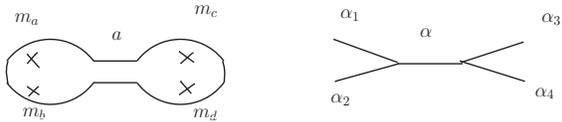}
\end{center}
\caption{On the left hand side is the Riemann surface associated
with the $SU(2)$ gauge theory in one particular S dual frame, the
Coulomb branch parameter is $a$. One the right hand side, we draw
the $s$ channel contribution to the four point function of Liouville
theory.}
\end{figure}

Now the crossing symmetry (\ref{1}) of CFT states that the
correlation function of different channels are related. When we
consider the gauge theory, the different channels mean different $S$
dual frames (see Figure 3). With the identification between the
partition function of gauge theory and correlation function of CFT,
we conclude that:

The partition function of this four dimensional SCFT in different S
dual frames are related as in formula $(\ref{1})$. Notice that the
gauge coupling is identified with the position of the unfixed
coordinate of the vertex operator, and the second identity in
$(\ref{1})$ relates theories with gauge couplings $q^{'}={1\over
q}$. So this identity implies that the partition function of one
strongly coupled theory is determined by another weakly coupled
theory as predicted by S-duality! We have determined the exact
relations between them from Liouville theory.

Next let's consider $N=4$ $SU(2)$ theory, then we are tempting to
identify the gauge theory partition function with the partition
function of Liouville theory, the Liouville partition function can
be calculated from (\ref{3}):
\begin{equation}
Z(\tau)=V_\phi{1\over2\pi\sqrt{\tau_2}|\eta(\tau)|^2}
\end{equation}
here $V_\phi$ is the zero mode contribution and is independent of
$\tau$, this is identified with the gauge theory partition function;
we also need to identify $\alpha=1+a$ for intermediate state and the
gauge coupling is ${\pi\over g_{YM}^2}=\tau_2$ which is consistent
with our previous identification. Comparing to (\ref{4}) with $M=2$,
we can see that the $U(1)$ part contribution to gauge theory
partition function is
\begin{equation}
Z^{U(1)}=C^{'}{1\over 2\pi\sqrt{\tau_2}|\eta(\tau)|^2}.
\end{equation}
This result can be generalized to $SU(N)$ case in which the
$A_{N-1}$ conformal Toda field theory is involved. The one-loop
partition function of $A_{N-1}$ theory is equivalent to $N-1$ free
scalars so the partition function has a factor ${1\over
2\pi\sqrt{\tau}|\eta(\tau)|^{2N-2}}$, comparing with (\ref{4}), the
$U(1)$ part contributes a factor ${1\over
2\pi\sqrt{\tau_2}|\eta(\tau)|^2}$.

\section{Conclusion}
In this note, we derive the transformation law for the partition
function of certain SU(2) SCFT under the S duality transformation
from Liouville theory point of view. It is easy to generalize to
higher rank gauge theory (for higher rank gauge theory, the CFT side
is the conformal Toda field theory \cite{toda}).

Consider a six dimensional $(0,2)$ theory compactified on a
punctured Riemann surface $\Sigma$, we get a four dimensional gauge
theory on a four manifold $M$; on the other hand, we can first
compactify six dimensional theory on a four manifold $M$ (with
possible singularities), we can get a CFT on $\Sigma$. It is
interesting to identify $\Sigma$ and $M$ and the corresponding gauge
theory and conformal field theory. We may also have the beautiful
correspondence between gauge theory and CFT.

Finally, Liouville theory plays a fundamental role in non-critical
string theory , and it has interesting application to
cosmology\cite{aben}. Any understanding of the dynamics of the
theory is interesting. The dynamical information of the theory is
encoded in the structure constant of the theory. This can be
calculated from the one-loop part of the partition function. It is
interesting to see what we can learn from gauge theory about the
property of Liouville theory, for instance: the spectrum, the
Seiberg bound...

\begin{acknowledgments}
This research was supported in part by the Mitchell-Heep Chair in High Energy Physics,
 and by the DOE grant DE-FG$03$-$95$-Er-$40917$(DVN).

\end{acknowledgments}

\end{document}